# A platform for *in situ* synthesis in a STEM

*Ondrej Dyck[1], Andrew R. Lupini[1], Stephen Jesse[1]*

[1] *Center for Nanophase Materials Sciences, Oak Ridge National Laboratory, Oak Ridge, TN*

**Abstract**

The engineering of quantum materials requires the development of tools able to address various synthesis and characterization challenges. These include the establishment and refinement of growth methods, material manipulation, and defect engineering. Material modification at the atomic level will be a key enabling factor for the engineering of quantum materials where desired phenomena are critically determined by local atomic structures. Successful use of scanning transmission electron microscopes (STEMs) for atomic scale material manipulation has opened the door for a transformed view of what can be accomplished using electron-beam-based strategies. However, serious obstacles exist on the pathway from possibility to practical reality. One such obstacle is the *in situ* delivery of atomized material in the STEM to the region of interest for further fabrication processes. Here, we present progress on this front with a view toward performing synthesis (deposition and growth) processes in a scanning transmission electron microscope. An *in situ* thermal deposition platform is presented, tested, and deposition and growth processes are demonstrated. In particular, we show that isolated Sn atoms can be evaporated from a filament and caught on the nearby sample, demonstrating atomized material delivery. This platform, and future variations, are envisioned to facilitate real-time atomic resolution imaging of growth processes and open new pathways toward atomic fabrication.

**Introduction**

Growth and manipulation of materials down to the atomic level are viewed as key enabling factors for the future of materials engineering, especially in the domain of quantum materials where atomic precision may be critical to provide conditions for desired quantum phenomena to emerge and persist. A priority research direction in the quantum materials domain includes the development of tools capable of addressing synthesis and characterization challenges in establishing methods of growth, manipulation, and the control of dopants and defects.[1] In this endeavor, *in situ* diagnostics and real-time adaptive feedback control are important capabilities that can accelerate understanding of the underlying physics and chemistry involved in material growth and the response to environmental influences.

In the last few years, the scanning transmission electron microscope (STEM) has been reconceptualized, transitioning from characterization-only to a characterization-plus-manipulation platform.[2–5] Numerous demonstrations have ranged from the creation of nanowires[6,7] to 1D chains of atoms,[8–12] sculpting,[13–17] molecule-by-molecule deposition,[18] the movement of single atoms,[19–23] the attachment of atoms,[24–28] and atomic patterning[29] and writing.[30] At the same time, feedback-controlled methods for material transformation, manipulation, and beam control in a STEM have also become increasingly refined, now leveraging artificial-intelligence (AI) based decision making.[31–38] Likewise, advances in bottom-up synthesis techniques have shown an impressive ability to tailor molecular structures with atomic precision, revealing junction states, spin centers, quantum spin chains, topologically induced energy bands, and spin splitting.[39–49] Characterization of these formed structures has received a critical advancement with the ability to effectively transfer them into high-resolution instruments, capable of examining the emergent properties.[44] This underscores the value in not simply forming the structures but also being able to examine them with atomic precision. What if, instead of transferring synthesized materials into such an instrument, we were able to directly synthesize and modify the materials within the microscope chamber and study them there?

In a recently published paper, we laid out a vision for merging of the fields of e-beam atomic manipulation and synthesis in a STEM, dubbed the "synthescope".[50] In this conceptualization, the synthesis processes would be performed inside the microscope vacuum chamber while being directly observed with atomic resolution in real time. Moreover, the material interaction with the focused e-beam would create a movable perturbation that would effectively function as an atomic-sized reaction chamber, within which relevant factors/variables that govern the behavior of the system (i.e. physical and chemical evolution) are vastly different from the surrounding material. This would enable the tuning of global parameters (e.g. temperature, optical excitation, partial pressures, etc.) while confining the action to the irradiated area. A pivotal concept, however, is the delivery of some external reactant as is done in molecular beam epitaxy (MBE), chemical vapor deposition (CVD), pulsed laser deposition (PLD), or similar techniques. But the reaction would be directed locally and top-down similar to e-beam induced deposition (EBID). Importantly, this differs from STEM-based EBID in that it is not confined to the dissociation of a precursor gas but also includes working with purified precursor materials.[29,30] This would enable multi-component synthesis processes combined with atomic-scale tailoring, real-time feedback control, and AI decision making.

Here, we present initial progress on this front, detailing the development of a self-contained (i.e. no major modifications to the existing microscope hardware) evaporative platform incorporated into the sample holder aimed at *in situ* delivery of precursor material to a sample in the STEM.

## *In Situ* Evaporation Platform

The Nion microscope company, in collaboration with Protochips[TM],[51–54] have developed a platform that is designed to enable electrical connection and operation of a variety of electrical chips (e-chips) upon which one's sample may be mounted to facilitate thermal and electrical experiments. This platform includes an electrical receiver cartridge that interfaces with the microscope hardware (Nion) and a removable insert that interfaces with the e-chips (Protochips). Here, we introduce the design of our own removable insert that interfaces with the existing receiver cartridge, as well as the e-chips, but provides an additional external (not incorporated into the e-chip) heater element. In this approach, the external heater element can be used for material evaporation and delivery to the sample. The temperature of the sample can be independently controlled by using the built-in heater of the e-chip. Filament-style heaters have been used in previous investigations,[55] however, here we do not aim to heat the sample but rather evaporate and deliver material to the sample while at the same time maintaining the functionality of existing e-chip platforms to expand the range of capabilities.

The design consists of two main components shown in Figure 1 labels (1) and (3). These are: (1) a printed circuit board (PCB) furnished with electrical contact pads to interface with the existing electrical receiver cartridge, an electrical contact block to interface with existing e-chips, and additional contact pads to facilitate the attachment of a resistive heater element, and (3) a metal base plate with locating pins and a central aperture to allow for the transmission of the e-beam. These two pieces are screwed together with the e-chip device/sample (2) positioned by the locating pins and held in place by the contact block, illustrated in Figure 1 label (4). This assembly slides into the Nion electrical receiver cartridge (5), interfacing with and held in position by the electrical contact springs in the cartridge. The full assembly is shown in Figure 1 label 6. This assembly is fully compatible with existing Nion microscope hardware and can be vacuum baked and loaded into the microscope using standard procedures. A photograph of the assembly on the bench is shown in Figure 1 label (7) with a magnified view of the heater filament. Note, the diagram of the Nion electrical cartridge is only representational and has been substantially simplified.

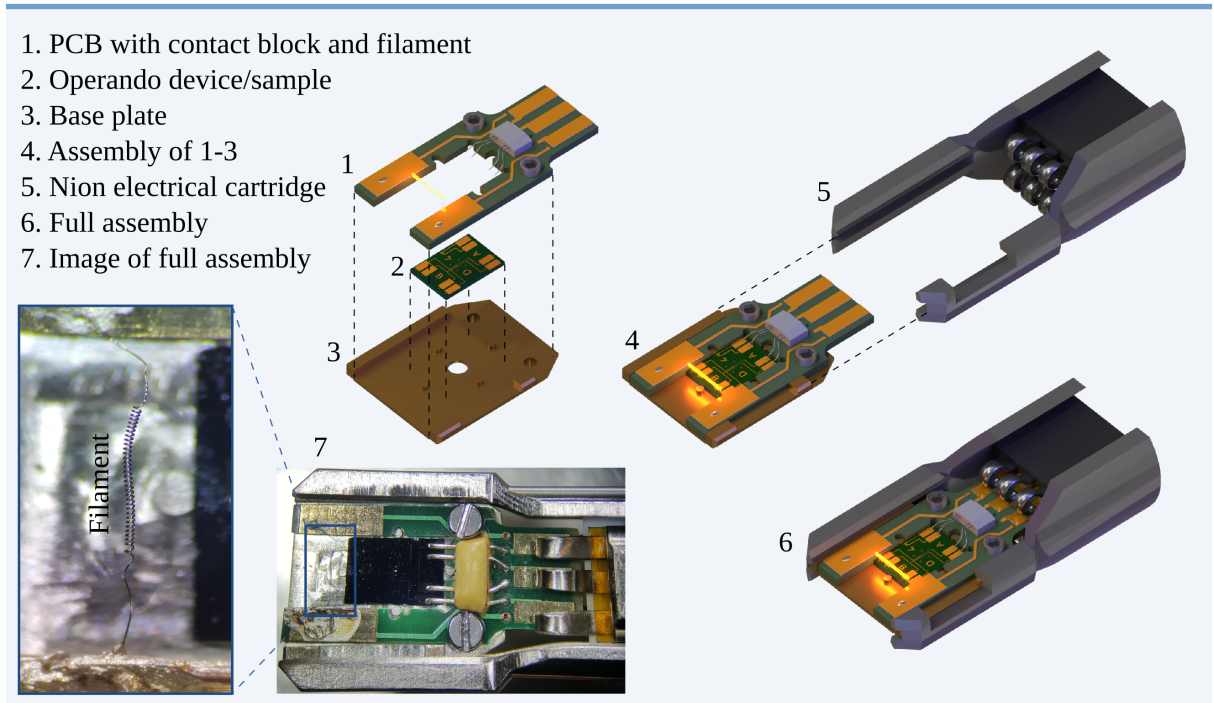

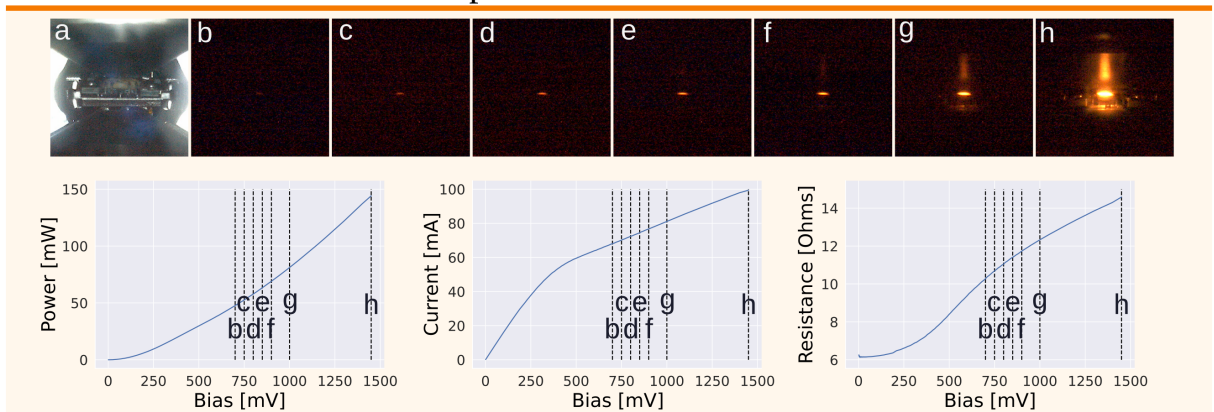

**Figure 1 Evaporation platform and trial run.** Top panel: diagram of the various evaporation platform components and how they assemble and interface with the existing hardware. An optical image of the assembly is shown in 7 with a zoomed in view of the evaporative filament. Bottom panel: Summary of an operational trial run. (a) shows the cartridge in the microscope pole piece. Images (b)-(h) show repeated images with the pole piece illumination turned off, acquired at various applied biases corresponding to the dotted vertical lines in the lower plots.

## Operational Trial Run

An operational trial run is summarized in the lower panel of Figure 1. In this trial run, the bare W filament was tested for *in situ* operation. Image (a) shows the camera view into the pole piece with the assembly inserted (camera illumination on). Images (b)-(h) show the same view with the camera illumination turned off and the filament at various applied voltages. The image labels (b)-(h) correspond to the dotted lines on the plots below which summarize the electrical response of the filament up to a maximum current of 100 mA. This corresponded to a power consumption of just under 150 mW. As can be observed in the images, the filament glows brightly at this power output.

**Evaporation Trial**

As a first test for evaporation, solder (composed of Sn, Bi, and Cu) was melted onto the filament surface. This was done by briefly heating the filament in atmosphere above the melting temperature of the solder and allowing enough time for the solder to wet the filament. This provided a low melting point material on the heater filament to evaporate. The test sample was a custom-fabricated e-chip with an electron transparent $SiN_x$ membrane at the center to allow for STEM imaging. The e-chip fabrication details are presented elsewhere.[56] However, for the purposes of this initial experiment, the functionality of the e-chip is unimportant. The e-chip was used simply to provide an electron transparent membrane upon which to evaporate and subsequently image any evaporated/condensed material.

The initial sample state is show in Figure 2(a). To protect the other vacuum chambers of the microscope, and particularly the gun vacuum, the vacuum valves to the objective chamber were closed, which prevented imaging during the evaporation process. After the evaporation was completed, the valves were reopened and the image shown in Figure 2(b) was acquired. Nanoparticles are observed covering the $SiN_x$ membrane indicating that solder had evaporated from the W filament and condensed onto the $SiN_x$ surface.

To accomplish the evaporation, a series of voltages were applied to the filament using a Keithley 2636B SourceMeter voltage source. The voltage source was operated manually and each voltage was only applied long enough to obtain a current reading (1-3 s). The voltage, current, and filament resistance readings are shown in Figure 2(f). The vacuum level of the objective chamber was recorded throughout this process and is shown in Figure 2(c). Vacuum spikes are marked with the corresponding power output of the filament. After each spike in the vacuum reading the experiment was paused to allow the vacuum to recover. No change in the vacuum reading was observed until the application of 600 mV (36 mW). The application of 700 mV (46 mW) resulted in the largest vacuum spike, followed by three approximately equal vacuum spikes for the application of 750-850 mV (51-61 mW). The most likely explanation for the decreased evaporation rate (lower vacuum spikes) for these higher voltage applications is that the majority of the volatile material evaporated from the filament during the application of the 700 mV (46 mW) which caused the largest vacuum spike, resulting in less remaining material to evaporate at these higher powers. This observation suggests that a more carefully controlled strategy may be needed to tune the rate of material supply. The image in Figure 2(b) was acquired at the time marked "Image Acquired" in Figure 2(c).

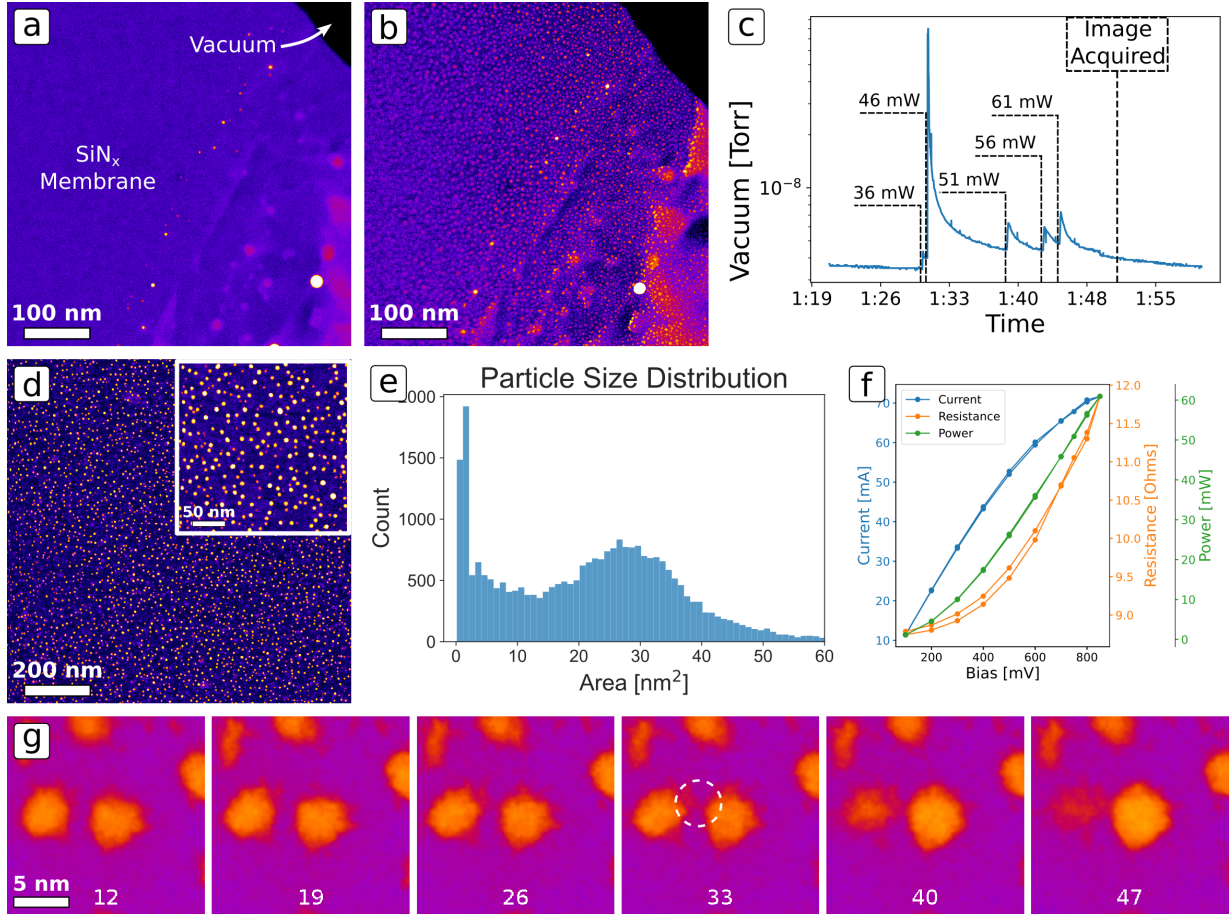

**Figure 2 Evaporative delivery of nanoparticles.** (a) Initial state of the $SiN_x$ membrane. (b) Final state of the $SiN_x$ membrane. (c) Recorded vacuum reading of the pole piece vacuum chamber through time. Filament power consumption is labeled as well as the time when the image in (b) was recorded. (d) Higher resolution images of the $SiN_x$ membrane with deposited nanoparticles. (e) Histogram of the size distribution of the nanoparticles. (f) Voltage, current, resistance, and power recorded during the evaporation process. (g) Example observation of nanoparticle agglomeration.

Figure 2(d) shows magnified views of the SiN$_x$ membrane with the nanoparticles more clearly visible. A computer vision approach to topological structural analysis of binary images,[57] implemented by OpenCV,[58] was used to find the nanoparticle contours and extract the particle cross-sectional areas. A histogram of the extracted particle areas is shown in Figure 2(e). We see that the areas follow a bimodal distribution with a peak around 30 nm$^2$ (~6 nm in diameter) and a second, sharper peak close to zero. The results from each analyzed image are presented in the SI and the code that reproduces these results can be found at https://github.com/ondrejdyck/A-platform-for-in-situ-synthesis-in-a-STEM.

The envisioned application for this platform is to deliver atomized materials to a sample and directly image synthesis processes in real time with atomic resolution. Thus, it is encouraging to observe this sharp peak representing very small nanoparticle sizes. In these images, we find that an analysis of a higher resolution image of the sample (shown in the SI) reveals a sharper and taller peak near zero indicating that most of the images acquired for this analysis were not acquired with high enough resolution to distinguish the smallest particles (which can be single atoms). This subsequent observation confirms the lower peak is real and not an artifact representing the spurious detection of noise as particles.

It seems likely that the larger nanoparticles were formed from the coalescence of smaller particles ultimately formed from the coalescence of individual atoms migrating along the SiN$_x$ surface. However, these dynamics were not observed during the imaging used for the particle size analysis. To investigate this hypothesis, the heater filament was later reactivated (800 mV) and the sequence of images shown in Figure 2(g) acquired, which clearly show the merging of two nanoparticles. This dataset is presented as a video in the SI and the numbers at the bottom of the images here, correspond to the video frame numbers. We conclude that the heat from the filament is sufficient to induce the surface migration and continued coarsening of the nanoparticles. This conclusion is consistent with the proposition that the larger particles are formed by aggregation rather than direct deposition.

**Atomized Deposition**

*Growth of 2D nanoplatelets*

The evaporation observed in Figure 2 occurred rapidly at an applied voltage of 700 mV (46 mW). Subsequent heating revealed that most of the solder evaporated during this initial heating. Furthermore, the total quantity of material deposited depends on the quantity available. Thus, to limit the amount of deposited material and investigate the smallest particles which form during the initial phase of deposition we used a W filament coated with 500 nm of Sn using e-beam evaporation. This approach for filament preparation provides a more controllable amount of relatively pure source material.

In the previous experiment, single atoms were somewhat difficult to image clearly through the $SiN_x$ membrane but can be seen on close inspection of Figure 2(g). Therefore in this experiment, a Protochips™ heating e-chip was used as the substrate to enable independent control over the sample temperature and a graphene bilayer was suspended across the e-chip apertures to enable clear imaging of any single atoms or atomic clusters deposited on the surface. After introduction into the microscope the heating e-chip was ramped to 1200 °C at a ramp rate of 1000 °C/ms, held for a few minutes, then ramped back to 500 °C at a ramp rate of 100 °C/s and left overnight to achieve thermal equilibrium. The initial heating has been shown to remove surface hydrocarbon contamination[59–61] and the sustained elevated temperature is conjectured to combat further hydrocarbon deposition that has been observed on what appears to be clean graphene.[62] The state of the sample after this procedure is shown in Figure 3(a) where we observe a majority of bilayer graphene suspended over an aperture in the e-chip with a number of wrinkles.

A magnified view of the bilayer graphene is shown in Figure 3(b), where the Moire pattern between layers is visible. The accelerating voltage of the microscope was 100 kV, which is above the knock-on energy for graphene. Consequently, we observe the gradual generation of point defects and defect clustering in the bilayer. Several examples are boxed in the figure with magnified views overlaid.

The evaporative procedure was performed similarly to the previous example, but with the vacuum valves remaining open to allow live imaging to monitor the sample for changes. The recorded vacuum levels are presented in Figure 3(d) and the voltage, current, resistance, and power of the filament are presented in Figure 3(e). At 90 mA evidence of deposition was observed and the image shown in Figure 3(c) was recorded, where bright particles can be seen attached to many step edges and in various locations on the graphene bilayer. Given the magnification level, it is possible that some deposition occurred prior to this but that it was not observed. Notably, the circled region indicates the location where previous imaging, as shown in Figure 3(b), was recorded and defects introduced.

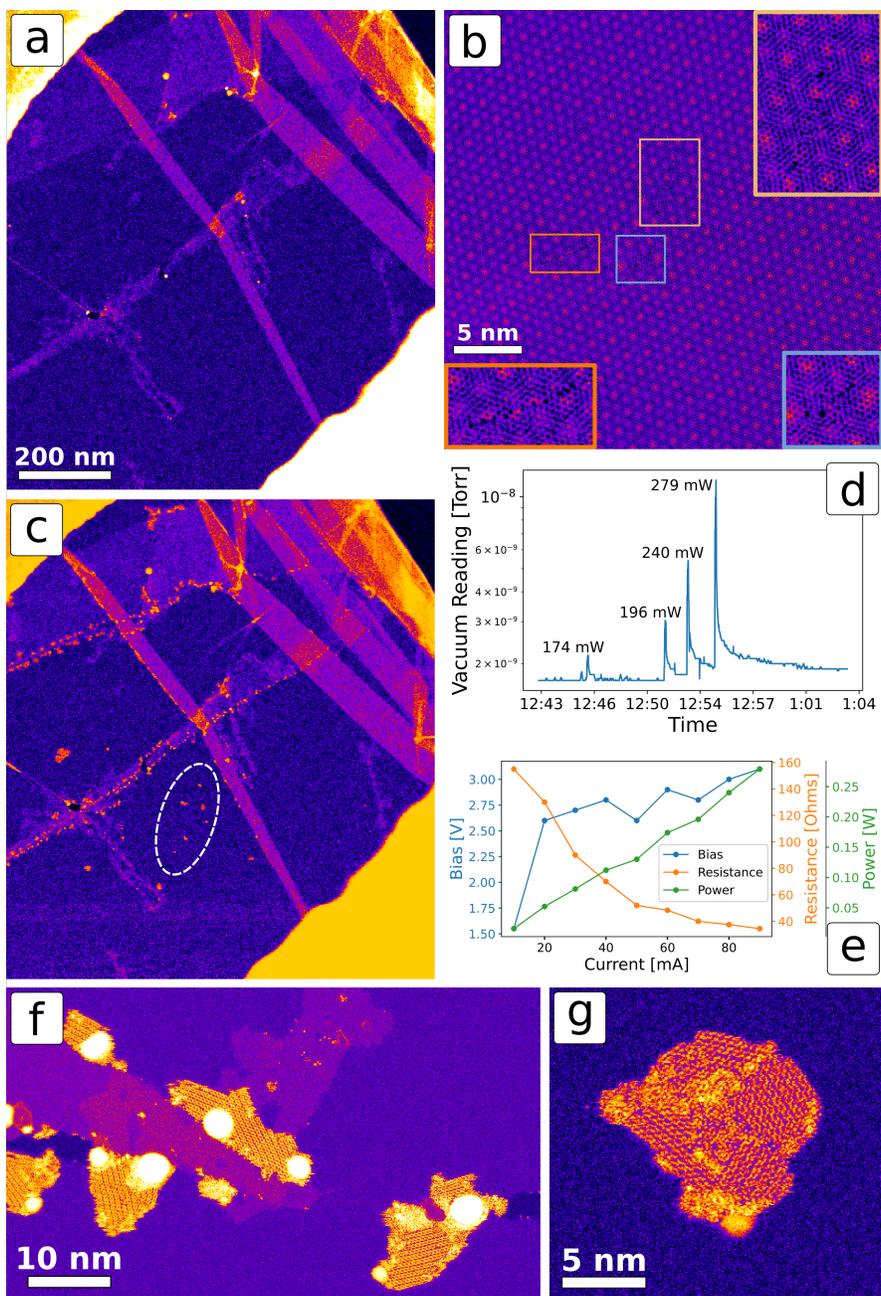

**Figure 3 *In situ* growth of nano-platelets.** (a) Initial sample overview at 500 °C. (b) Magnified view of bilayer graphene with e-beam generated defects, boxed. (c) Overview of sample after evaporation. Dotted oval marks the region that was damaged by the e-beam. (d) Vacuum reading through time. Vacuum spikes are annotated with the current that was applied to the filament. (e) Bias, resistance, and power consumed by the filament as a function of applied current. (f) Representative image of nano-platelets growing at step edges. (g) Representative image of nano-platelet growing away from step edges on the graphene bilayer.

Figure 3(f) and (g) show magnified images of these particles. They appear to be composed of Sn nanoparticles and 2D nanoplatelets attached primarily to step edges. Some nanoplatelets appear to have nucleated in the middle of the graphene bilayer, most likely at defect locations. These nanoplatelets were found to be sensitive to the e-beam and were quickly destroyed. A video of this damage is supplied in the SI. To assess the composition of the destroyed nanoplatelets electron energy loss spectroscopy was performed (shown in the SI), revealing only C and Sn, which suggests that the nanoplatelets are 2D Sn. However, it is possible that they form as an oxide of Sn and that radiation damage releases the oxygen during the exposure as has been reported for Cr[28] and other oxides.[63] The precise determination of the chemistry and structure is beyond the scope of the present paper.

It is unlikely that these nanoplatelets formed upon evaporation and were deposited fully formed on the graphene surface. More likely, single atoms or few atom clusters were evaporated and deposited on the graphene. Given that the majority of the graphene surface is atomically perfect, with few binding sites, and that the temperature of the surface was maintained at 500 °C, these atoms were then able to migrate and coalesce into the observed 2D nanoplatelets. If this hypothesis is correct, we should also find evidence of single atoms attached elsewhere on the sample that have been prevented from coalescing with other atoms.

*Further evidence of single atoms*

Examination of other areas of the sample provides ample evidence for the existence of single isolated atoms. Figure 4(a) shows a bilayer/trilayer step edge harboring many single atoms. Figure 4(b) shows single layer/bilayer and single layer/trilayer step edges harboring many single atoms. We note again that few single atoms are attached to the pristine layers, but instead are primarily found at the step edges. An exception to this is shown in Figure 4(c) where numerous single atoms are attached to a graphene layer that has an irregular shape with holes in it. This morphology is consistent with that observed in the graphitization of surface contaminants on heating,[59] meaning that this type of layer was likely formed during the initial heating of the e-chip. Given the irregular shape, the presence of holes, and the additional non-planar layers that can be observed in Figure 4(c) it is likely that these graphitized layers are imperfect, containing many more attachment points for the Sn atoms than the underlying bilayer graphene.

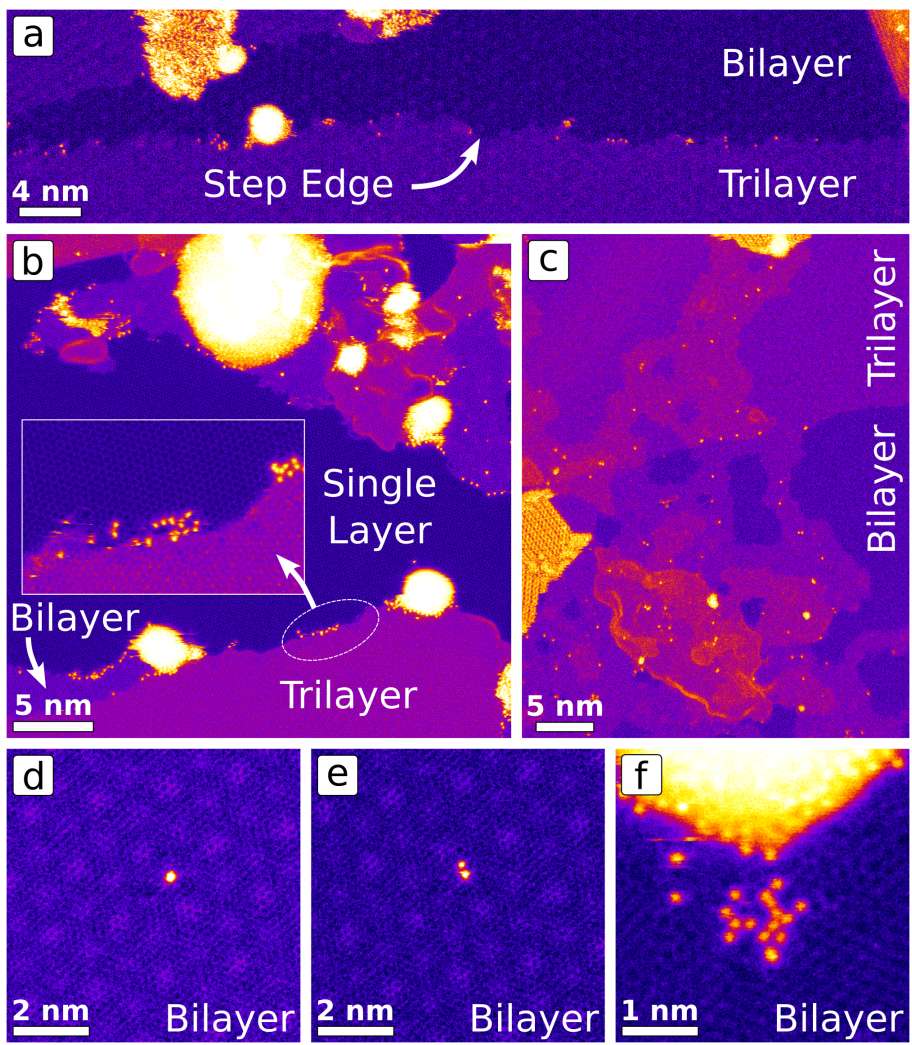

**Figure 4 Examination of single atom attachment sites.** (a) Bilayer/trilayer step edge harboring single atoms. (b) Single layer/trilayer and single layer/bilayer step edges harboring single atoms. (c) Single atoms attached to graphetized contamination layer. (d) A single atom attached to a bilayer. (e) An atomic dimer attached to a bilayer. (f) A cluster of atoms attached to a bilayer induced by e-beam irradiation and the generation of attachment sites.

If it is true that the Sn atoms are attaching to defects in the graphitized contamination layer, we should be able to find some single atoms attached to the graphene bilayer at defect locations. Two examples are shown in Figure 4(d) and (e). While it is not directly clear from the images that these atoms are attached to defect sites, this is most likely the case since the majority of the bilayer is free of the presence of single atoms. Moreover, the energy imparted to the atoms by the e-beam during the imaging process can promote spontaneous diffusion making adatoms difficult to image if they are not strongly bound to defect sites because of the sequential acquisition process. To illustrate the attachment of Sn atoms to defect sites more conclusively, the e-beam was scanned across a Sn nanoparticle and the graphene bilayer. As defects were introduced into the graphene, Sn atoms readily attached to the bilayer, shown in Figure 4(f).

With this evidence, we conclude that the Sn nanoplatelets were formed through the spontaneous diffusion of deposited Sn atoms along the surface of the graphene. Detailed theoretical modeling of this synthesis process and a structural analysis of the formed nanoplatelets is underway and will be presented elsewhere. Attachment points, consisting of graphene step edges and defects, form the nucleation sites for these nanoplatelets which may provide a pathway toward patterned synthesis strategies.

**Conclusion**

Here, we have presented the design concept and proof of operation of an *in situ* STEM evaporation platform. This platform is envisioned as an enabling step toward a generalized synthesis microscope ("synthescope").[50] We have successfully evaporated and deposited solder nanoparticles on a $SiN_x$ membrane and have refined this process to single-atom delivery by using a thinner coating of more chemically pure Sn on the filament. Sn nanoplatelets were synthesized *in situ* at graphene step edges and at damaged regions of the graphene bilayer. We found compelling evidence that the nanoplatelets were formed through the surface diffusion of Sn atoms.

Targets for future improvements include a greater degree of control over the filament temperature, possibly using a feedback signal to control the power consumption. In addition, substantial thermal drift during the filament operation was found to be a drawback for the current implementation. As one of the goals of performing this process in a STEM is to allow atomic resolution imaging and manipulation of the synthesized structures while they are forming, mitigating this thermal drift represents a significant avenue for future improvement.

The advancement demonstrated here may provide a mechanism to initiate patterned synthesis processes that combine top-down control with spontaneous bottom-up growth.[29] For example, one might create a pattern of defect sites (e.g. arrays, lines, spirals, etc.) after which an evaporation/nucleation/growth procedure positions nanoplatelets at each defect location. Such a strategy would combine the atomic scale precision of STEM-based imaging and manipulation with the speed of more conventional synthesis processes. Given the range of two-dimensional materials that can now be produced and the range of elements that might be suitable for this deposition process, the array of growth possibilities is immense.

**Methods**

*Scanning Transmission Electron Microscope*

The described experiments were performed in a Nion UltraSTEM 200. The initial evaporation presented in Figure 2 was performed with an accelerating voltage of 200 kV. The high angle annular dark field (HAADF) detector was used for imaging. The second evaporation, presented in Figure 3 and Figure 4, was performed with an accelerating voltage of 100 kV and current of 16 pA. The medium angle annular dark field (MAADF) detector was used for imaging. Electron energy loss spectroscopy (EELS) was acquired using a Gatan Enfinium spectrometer, using a nominal convergence angle of 30 mrad and collection angle of 32 mrad.

*Heater Platform*

The base plate for the heater platform is a custom design and was fabricated in-house from Ti metal. The PCB and contact block were fabricated by Wentworth Laboratories based on provided designs. The W filaments were extracted from new incandescent Christmas lights and manually soldered to the PCB.

**Acknowledgement**

This work was supported by the U.S. Department of Energy, Office of Science, Basic Energy Sciences, Materials Sciences and Engineering Division (O.D. A.R.L., S.J.), and was performed at the Center for Nanophase Materials Sciences (CNMS), a U.S. Department of Energy, Office of Science User Facility.

# References

(1)     Broholm, C.; Fisher, I.; Moore, J.; Murnane, M.; Moreo, A.; Tranquada, J.; Basov, D.; Freericks, J.; Aronson, M.; MacDonald, A.; Fradkin, E.; Yacoby, A.; Samarth, N.; Stemmer, S.; Horton, L.; Horwitz, J.; Davenport, J.; Graf, M.; Krause, J.; Pechan, M.; Perry, K.; Rhyne, J.; Schwartz, A.; Thiyagarajan, T.; Yarris, L.; Runkles, K. *Basic Research Needs Workshop on Quantum Materials for Energy Relevant Technology*; USDOE Office of Science (SC) (United States), 2016. https://doi.org/10.2172/1616509.
(2)     Kalinin, S. V.; Borisevich, A.; Jesse, S. Fire up the Atom Forge. *Nature* **2016**, *539* (7630), 485–487.
(3)     Dyck, O.; Jesse, S.; Kalinin, S. V. A Self-Driving Microscope and the Atomic Forge. *MRS Bull.* **2019**, *44* (9), 669–670. https://doi.org/10.1557/mrs.2019.211.
(4)     Dyck, O.; Ziatdinov, M.; Lingerfelt, D. B.; Unocic, R. R.; Hudak, B. M.; Lupini, A. R.; Jesse, S.; Kalinin, S. V. Atom-by-Atom Fabrication with Electron Beams. *Nat. Rev. Mater.* **2019**, *4* (7), 497–507. https://doi.org/10.1038/s41578-019-0118-z.
(5)     Susi, T.; Kepaptsoglou, D.; Lin, Y.-C.; Ramasse, Q. M.; Meyer, J. C.; Suenaga, K.; Kotakoski, J. Towards Atomically Precise Manipulation of 2D Nanostructures in the Electron Microscope. *2D Materials* **2017**, *4* (4), 042004.
(6)     Lin, J.; Cretu, O.; Zhou, W.; Suenaga, K.; Prasai, D.; Bolotin, K. I.; Cuong, N. T.; Otani, M.; Okada, S.; Lupini, A. R.; Idrobo, J.-C.; Caudel, D.; Burger, A.; Ghimire, N. J.; Yan, J.; Mandrus, D. G.; Pennycook, S. J.; Pantelides, S. T. Flexible Metallic Nanowires with Self-Adaptive Contacts to Semiconducting Transition-Metal Dichalcogenide Monolayers. *Nat. Nanotechnol.* **2014**, *9*, 436. https://doi.org/10.1038/nnano.2014.81 https://www.nature.com/articles/nnano.2014.81#supplementary-information.
(7)     Lin, J.; Zhang, Y.; Zhou, W.; Pantelides, S. T. Structural Flexibility and Alloying in Ultrathin Transition-Metal Chalcogenide Nanowires. *ACS Nano* **2016**, *10* (2), 2782–2790. https://doi.org/10.1021/acsnano.5b07888.
(8)     Andrey Chuvilin; Jannik C. Meyer; Gerardo Algara-Siller; Ute Kaiser. From Graphene Constrictions to Single Carbon Chains. *New Journal of Physics* **2009**, *11* (8), 083019.
(9)     Jin, C.; Lan, H.; Peng, L.; Suenaga, K.; Iijima, S. Deriving Carbon Atomic Chains from Graphene. *Phys. Rev. Lett.* **2009**, *102* (20), 205501. https://doi.org/10.1103/PhysRevLett.102.205501.
(10)    Cretu, O.; Komsa, H.-P.; Lehtinen, O.; Algara-Siller, G.; Kaiser, U.; Suenaga, K.; Krasheninnikov, A. V. Experimental Observation of Boron Nitride Chains. *ACS Nano* **2014**, *8* (12), 11950–11957. https://doi.org/10.1021/nn5046147.
(11)    Lin, Y.-C.; Morishita, S.; Koshino, M.; Yeh, C.-H.; Teng, P.-Y.; Chiu, P.-W.; Sawada, H.; Suenaga, K. Unexpected Huge Dimerization Ratio in One-Dimensional Carbon Atomic Chains. *Nano Lett.* **2017**, *17* (1), 494–500. https://doi.org/10.1021/acs.nanolett.6b04534.
(12)    Xiao, Z.; Qiao, J.; Lu, W.; Ye, G.; Chen, X.; Zhang, Z.; Ji, W.; Li, J.; Jin, C. Deriving Phosphorus Atomic Chains from Few-Layer Black Phosphorus. *Nano Res.* **2017**, *10* (7), 2519–2526. https://doi.org/10.1007/s12274-017-1456-z.
(13)    Clark, N.; Lewis, E. A.; Haigh, S. J.; Vijayaraghavan, A. Nanometre Electron Beam Sculpting of Suspended Graphene and Hexagonal Boron Nitride Heterostructures. *2D Materials* **2019**, *6* (2), 025032. https://doi.org/10.1088/2053-1583/ab09a0.
(14)    Fischbein, M. D.; Drndić, M. Sub-10 Nm Device Fabrication in a Transmission Electron Microscope. *Nano Lett.* **2007**, *7* (5), 1329–1337. https://doi.org/10.1021/nl0703626.
(15)    Michael D. Fischbein; Marija Drndić. Electron Beam Nanosculpting of Suspended Graphene Sheets. *Appl. Phys. Lett.* **2008**, *93* (11), 113107. https://doi.org/10.1063/1.2980518.
(16)    Qi, Z. J.; Rodríguez-Manzo, J. A.; Botello-Méndez, A. R.; Hong, S. J.; Stach, E. A.; Park, Y. W.; Charlier, J.-C.; Drndić, M.; Johnson, A. T. C. Correlating Atomic Structure and Transport in Suspended Graphene Nanoribbons. *Nano Lett.* **2014**, *14* (8), 4238–4244. https://doi.org/10.1021/nl501872x.
(17)    Xu, Q.; Wu, M.-Y.; Schneider, G. F.; Houben, L.; Malladi, S. K.; Dekker, C.; Yucelen, E.; Dunin-Borkowski, R. E.; Zandbergen, H. W. Controllable Atomic Scale Patterning of Freestanding Monolayer Graphene at Elevated Temperature. *ACS Nano* **2013**, *7* (2), 1566–1572. https://doi.org/10.1021/nn3053582.
(18)    van Dorp, W. F.; Zhang, X.; Feringa, B. L.; Hansen, T. W.; Wagner, J. B.; De Hosson, J. T. M. Molecule-by-Molecule Writing Using a Focused Electron Beam. *ACS Nano* **2012**, *6* (11), 10076–10081. https://doi.org/10.1021/nn303793w.
(19)    Susi, T.; Kotakoski, J.; Kepaptsoglou, D.; Mangler, C.; Lovejoy, T. C.; Krivanek, O. L.; Zan, R.; Bangert, U.; Ayala, P.; Meyer, J. C.; Ramasse, Q. Silicon-Carbon Bond Inversions Driven by 60-KeV Electrons in Graphene. *Phys. Rev. Lett.* **2014**, *113* (11), 115501.

(20) Susi, T.; Meyer, J. C.; Kotakoski, J. Manipulating Low-Dimensional Materials down to the Level of Single Atoms with Electron Irradiation. *Ultramicroscopy* **2017**, *180*, 163–172. http://dx.doi.org/10.1016/j.ultramic.2017.03.005.

(21) Mustonen, K.; Markevich, A.; Tripathi, M.; Inani, H.; Ding, E.-X.; Hussain, A.; Mangler, C.; Kauppinen, E. I.; Kotakoski, J.; Susi, T. Electron-Beam Manipulation of Silicon Impurities in Single-Walled Carbon Nanotubes. *Advanced Functional Materials* **2019**, *29* (52), 1901327. https://doi.org/10.1002/adfm.201901327.

(22) Tripathi, M.; Mittelberger, A.; Pike, N. A.; Mangler, C.; Meyer, J. C.; Verstraete, M. J.; Kotakoski, J.; Susi, T. Electron-Beam Manipulation of Silicon Dopants in Graphene. *Nano Lett.* **2018**, *18* (8), 5319–5323. https://doi.org/10.1021/acs.nanolett.8b02406.

(23) Hudak, B. M.; Song, J.; Sims, H.; Troparevsky, M. C.; Humble, T. S.; Pantelides, S. T.; Snijders, P. C.; Lupini, A. R. Directed Atom-by-Atom Assembly of Dopants in Silicon. *ACS Nano* **2018**, *12* (6), 5873–5879. https://doi.org/10.1021/acsnano.8b02001.

(24) Dyck, O.; Kim, S.; Kalinin, S. V.; Jesse, S. Placing Single Atoms in Graphene with a Scanning Transmission Electron Microscope. *Appl. Phys. Lett.* **2017**, *111* (11), 113104. https://doi.org/10.1063/1.4998599.

(25) Dyck, O.; Kim, S.; Jimenez-Izal, E.; Alexandrova, A. N.; Kalinin, S. V.; Jesse, S. Building Structures Atom by Atom via Electron Beam Manipulation. *Small* **2018**, *14* (38), 1801771. https://doi.org/doi:10.1002/smll.201801771.

(26) Dyck, O.; Zhang, L.; Yoon, M.; Swett, J. L.; Hensley, D.; Zhang, C.; Rack, P. D.; Fowlkes, J. D.; Lupini, A. R.; Jesse, S. Doping Transition-Metal Atoms in Graphene for Atomic-Scale Tailoring of Electronic, Magnetic, and Quantum Topological Properties. *Carbon* **2021**, *173*, 205–214. https://doi.org/10.1016/j.carbon.2020.11.015.

(27) Dyck, O.; Zhang, C.; Rack, P. D.; Fowlkes, J. D.; Sumpter, B.; Lupini, A. R.; Kalinin, S. V.; Jesse, S. Electron-Beam Introduction of Heteroatomic Pt–Si Structures in Graphene. *Carbon* **2020**, *161*, 750–757. https://doi.org/10.1016/j.carbon.2020.01.042.

(28) Dyck, O.; Yoon, M.; Zhang, L.; Lupini, A. R.; Swett, J. L.; Jesse, S. Doping of Cr in Graphene Using Electron Beam Manipulation for Functional Defect Engineering. *ACS Appl. Nano Mater.* **2020**, *3* (11), 10855–10863. https://doi.org/10.1021/acsanm.0c02118.

(29) Dyck, O.; Yeom, S.; Lupini, A. R.; Swett, J. L.; Hensley, D.; Yoon, M.; Jesse, S. Top-down Fabrication of Atomic Patterns in Twisted Bilayer Graphene. arXiv January 4, 2023. https://doi.org/10.48550/arXiv.2301.01674.

(30) Dyck, O.; Lupini, A. R.; Jesse, S. Direct-Writing Atom-by-Atom. arXiv January 6, 2023. https://doi.org/10.48550/arXiv.2301.02743.

(31) Sang, X.; Lupini, A. R.; Unocic, R. R.; Chi, M.; Borisevich, A. Y.; Kalinin, S. V.; Endeve, E.; Archibald, R. K.; Jesse, S. Dynamic Scan Control in STEM: Spiral Scans. *Adv. Struct. Chem. Imaging* **2016**, *2* (1), 6. https://doi.org/10.1186/s40679-016-0020-3.

(32) Sang, X.; Lupini, A. R.; Ding, J.; Kalinin, S. V.; Jesse, S.; Unocic, R. R. Precision Controlled Atomic Resolution Scanning Transmission Electron Microscopy Using Spiral Scan Pathways. *Sci. Rep.* **2017**, *7* (1), 43585. https://doi.org/10.1038/srep43585.

(33) Jesse, S.; He, Q.; Lupini, A. R.; Leonard, D. N.; Oxley, M. P.; Ovchinnikov, O.; Unocic, R. R.; Tselev, A.; Fuentes-Cabrera, M.; Sumpter, B. G.; Pennycook, S. J.; Kalinin, S. V.; Borisevich, A. Y. Atomic-Level Sculpting of Crystalline Oxides: Toward Bulk Nanofabrication with Single Atomic Plane Precision. *Small* **2015**, *11* (44), 5895–5900. https://doi.org/10.1002/smll.201502048.

(34) Roccapriore, K. M.; Zou, Q.; Zhang, L.; Xue, R.; Yan, J.; Ziatdinov, M.; Fu, M.; Mandrus, D. G.; Yoon, M.; Sumpter, B. G.; Gai, Z.; Kalinin, S. V. Revealing the Chemical Bonding in Adatom Arrays via Machine Learning of Hyperspectral Scanning Tunneling Spectroscopy Data. *ACS Nano* **2021**, *15* (7), 11806–11816. https://doi.org/10.1021/acsnano.1c02902.

(35) Roccapriore, K. M.; Dyck, O.; Oxley, M. P.; Ziatdinov, M.; Kalinin, S. V. Automated Experiment in 4D-STEM: Exploring Emergent Physics and Structural Behaviors. *ACS Nano* **2022**, *16* (5), 7605–7614. https://doi.org/10.1021/acsnano.1c11118.

(36) Roccapriore, K. M.; Cho, S.-H.; Lupini, A. R.; Milliron, D. J.; Kalinin, S. V. Sculpting the Plasmonic Responses of Nanoparticles by Directed Electron Beam Irradiation. *Small* **2022**, *18* (1), 2105099. https://doi.org/10.1002/smll.202105099.

(37) Roccapriore, K. M.; Boebinger, M. G.; Dyck, O.; Ghosh, A.; Unocic, R. R.; Kalinin, S. V.; Ziatdinov, M. Probing Electron Beam Induced Transformations on a Single-Defect Level via Automated Scanning Transmission Electron Microscopy. *ACS Nano* **2022**, *16* (10), 17116–17127. https://doi.org/10.1021/acsnano.2c07451.

(38) Creange, N.; Dyck, O.; Vasudevan, R. K.; Ziatdinov, M.; Kalinin, S. V. Towards Automating Structural Discovery in Scanning Transmission Electron Microscopy. *Mach. Learn.: Sci. Technol.* **2022**, *3* (1), 015024. https://doi.org/10.1088/2632-2153/ac3844.

(39) Cao, T.; Zhao, F.; Louie, S. G. Topological Phases in Graphene Nanoribbons: Junction States, Spin Centers, and Quantum Spin Chains. *Phys. Rev. Lett.* **2017**, *119* (7), 076401. https://doi.org/10.1103/PhysRevLett.119.076401.

(40) Jiang, J.; Louie, S. G. Topology Classification Using Chiral Symmetry and Spin Correlations in Graphene Nanoribbons. *Nano Lett.* **2021**, *21* (1), 197–202. https://doi.org/10.1021/acs.nanolett.0c03503.

(41) Rizzo, D. J.; Veber, G.; Cao, T.; Bronner, C.; Chen, T.; Zhao, F.; Rodriguez, H.; Louie, S. G.; Crommie, M. F.; Fischer, F. R. Topological Band Engineering of Graphene Nanoribbons. *Nature* **2018**, *560* (7717), 204–208. https://doi.org/10.1038/s41586-018-0376-8.

(42) Gröning, O.; Wang, S.; Yao, X.; Pignedoli, C. A.; Borin Barin, G.; Daniels, C.; Cupo, A.; Meunier, V.; Feng, X.; Narita, A.; Müllen, K.; Ruffieux, P.; Fasel, R. Engineering of Robust Topological Quantum Phases in Graphene Nanoribbons. *Nature* **2018**, *560* (7717), 209–213. https://doi.org/10.1038/s41586-018-0375-9.

(43) Rizzo, D. J.; Veber, G.; Jiang, J.; McCurdy, R.; Cao, T.; Bronner, C.; Chen, T.; Louie, S. G.; Fischer, F. R.; Crommie, M. F. Inducing Metallicity in Graphene Nanoribbons via Zero-Mode Superlattices. *Science* **2020**, *369* (6511), 1597–1603. https://doi.org/10.1126/science.aay3588.

(44) McCurdy, R. D.; Jacobse, P. H.; Piskun, I.; Veber, G. C.; Rizzo, D. J.; Zuzak, R.; Mutlu, Z.; Bokor, J.; Crommie, M. F.; Fischer, F. R. Synergetic Bottom-Up Synthesis of Graphene Nanoribbons by Matrix-Assisted Direct Transfer. *J. Am. Chem. Soc.* **2021**, *143* (11), 4174–4178. https://doi.org/10.1021/jacs.1c01355.

(45) Blackwell, R. E.; Zhao, F.; Brooks, E.; Zhu, J.; Piskun, I.; Wang, S.; Delgado, A.; Lee, Y.-L.; Louie, S. G.; Fischer, F. R. Spin Splitting of Dopant Edge State in Magnetic Zigzag Graphene Nanoribbons. *Nature* **2021**, *600* (7890), 647–652. https://doi.org/10.1038/s41586-021-04201-y.

(46) Rogers, C.; Chen, C.; Pedramrazi, Z.; Omrani, A. A.; Tsai, H.-Z.; Jung, H. S.; Lin, S.; Crommie, M. F.; Fischer, F. R. Closing the Nanographene Gap: Surface-Assisted Synthesis of Peripentacene from 6,6′-Bipentacene Precursors. *Angewandte Chemie* **2015**, *127* (50), 15358–15361. https://doi.org/10.1002/ange.201507104.

(47) Marangoni, T.; Haberer, D.; Rizzo, D. J.; Cloke, R. R.; Fischer, F. R. Heterostructures through Divergent Edge Reconstruction in Nitrogen-Doped Segmented Graphene Nanoribbons. *Chemistry – A European Journal* **2016**, *22* (37), 13037–13040. https://doi.org/10.1002/chem.201603497.

(48) Perkins, W.; Fischer, F. R. Inserting Porphyrin Quantum Dots in Bottom-Up Synthesized Graphene Nanoribbons. *Chemistry – A European Journal* **2017**, *23* (70), 17687–17691. https://doi.org/10.1002/chem.201705252.

(49) Jacobse, P. H.; McCurdy, R. D.; Jiang, J.; Rizzo, D. J.; Veber, G.; Butler, P.; Zuzak, R.; Louie, S. G.; Fischer, F. R.; Crommie, M. F. Bottom-up Assembly of Nanoporous Graphene with Emergent Electronic States. *J. Am. Chem. Soc.* **2020**, *142* (31), 13507–13514. https://doi.org/10.1021/jacs.0c05235.

(50) Dyck, O.; Lupini, A. R.; Jesse, S. The Synthescope: A Vision for Combining Synthesis with Atomic Fabrication. arXiv February 16, 2023. https://doi.org/10.48550/arXiv.2302.08539.

(51) Sang, X.; Xie, Y.; Yilmaz, D. E.; Lotfi, R.; Alhabeb, M.; Ostadhossein, A.; Anasori, B.; Sun, W.; Li, X.; Xiao, K.; Kent, P. R. C.; van Duin, A. C. T.; Gogotsi, Y.; Unocic, R. R. In Situ Atomistic Insight into the Growth Mechanisms of Single Layer 2D Transition Metal Carbides. *Nat. Commun.* **2018**, *9* (1), 2266. https://doi.org/10.1038/s41467-018-04610-0.

(52) Sang, X.; Li, X.; Zhao, W.; Dong, J.; Rouleau, C. M.; Geohegan, D. B.; Ding, F.; Xiao, K.; Unocic, R. R. In Situ Edge Engineering in Two-Dimensional Transition Metal Dichalcogenides. *Nat. Commun.* **2018**, *9* (1), 2051. https://doi.org/10.1038/s41467-018-04435-x.

(53) Sang, X.; Li, X.; Puretzky, A. A.; Geohegan, D. B.; Xiao, K.; Unocic, R. R. Atomic Insight into Thermolysis-Driven Growth of 2D MoS2. *Advanced Functional Materials* **2019**, *29* (52), 1902149. https://doi.org/10.1002/adfm.201902149.

(54) Idrobo, J. C.; Lupini, A. R.; Feng, T.; Unocic, R. R.; Walden, F. S.; Gardiner, D. S.; Lovejoy, T. C.; Dellby, N.; Pantelides, S. T.; Krivanek, O. L. Temperature Measurement by a Nanoscale Electron Probe Using Energy Gain and Loss Spectroscopy. *Phys. Rev. Lett.* **2018**, *120* (9), 095901. https://doi.org/10.1103/PhysRevLett.120.095901.

(55) Kamino, T.; Saka, H. A Newly Developed High Resolution Hot Stage and Its Application to Materials Characterization. *Microsc. Microanal. Microstruct.* **1993**, *4* (2–3), 127–135. https://doi.org/10.1051/mmm:0199300402-3012700.

(56) Dyck, O.; Swett, J. L.; Evangeli, C.; Lupini, A. R.; Mol, J. A.; Jesse, S. Mapping Conductance and Switching Behavior of Graphene Devices In Situ. *Small Methods* **2022**, *6* (3), 2101245. https://doi.org/10.1002/smtd.202101245.

(57) Suzuki, S.; be, K. Topological Structural Analysis of Digitized Binary Images by Border Following. *Computer Vision, Graphics, and Image Processing* **1985**, *30* (1), 32–46. https://doi.org/10.1016/0734-189X(85)90016-7.

(58) Bradski, G. The OpenCV Library. *Dr. Dobb's Journal: Software Tools for the Professional Programmer*

**2000**, *25* (11).
(59)     Dyck, O.; Kim, S.; Kalinin, S. V.; Jesse, S. Mitigating E-Beam-Induced Hydrocarbon Deposition on Graphene for Atomic-Scale Scanning Transmission Electron Microscopy Studies. *Journal of Vacuum Science & Technology, B: Nanotechnology & Microelectronics: Materials, Processing, Measurement, & Phenomena* **2018**, *36* (1), 011801. https://doi.org/10.1116/1.5003034.
(60)     Tripathi, M.; Mittelberger, A.; Mustonen, K.; Mangler, C.; Kotakoski, J.; Meyer, J. C.; Susi, T. Cleaning Graphene: Comparing Heat Treatments in Air and in Vacuum. *physica status solidi (RRL) – Rapid Research Letters* **2017**, *11* (8), 1700124-n/a. https://doi.org/10.1002/pssr.201700124.
(61)     Lin, Y.-C.; Lu, C.-C.; Yeh, C.-H.; Jin, C.; Suenaga, K.; Chiu, P.-W. Graphene Annealing: How Clean Can It Be? *Nano Lett.* **2012**, *12* (1), 414–419. https://doi.org/10.1021/nl203733r.
(62)     Dyck, O.; Lupini, A. R.; Rack, P. D.; Fowlkes, J.; Jesse, S. Controlling Hydrocarbon Transport and Electron Beam Induced Deposition on Single Layer Graphene: Toward Atomic Scale Synthesis in the Scanning Transmission Electron Microscope. *Nano Select* **2022**, *3* (3), 643–654. https://doi.org/10.1002/nano.202100188.
(63)     Nan Jiang. Electron Beam Damage in Oxides: A Review. *Rep. Prog. Phys.* **2016**, *79* (1), 016501.

**Supplemental Information**

Figure S1 shows the particle analysis summary histogram for each image analyzed. Image 'SuperScan (HAADF) 44' was not included in the summary statistics presented in the main text because a different threshold was needed in the analysis process due to the presence of the dark vacuum signal. A different threshold may alter the detected particle sizes (although the distribution does not look substantially different from the rest) and so was omitted. This dataset is included here for completeness. Image 'SuperScan (HAADF) 41' was acquired at a higher pixel resolution (higher magnification) than the other images. We observe that the distribution of particle sizes is increased near zero, indicating a substantial number of very small particles are not resolved in the other images. The analysis code and images can be found at https://github.com/ondrejdyck/A-platform-for-in-situ-synthesis-in-a-STEM.

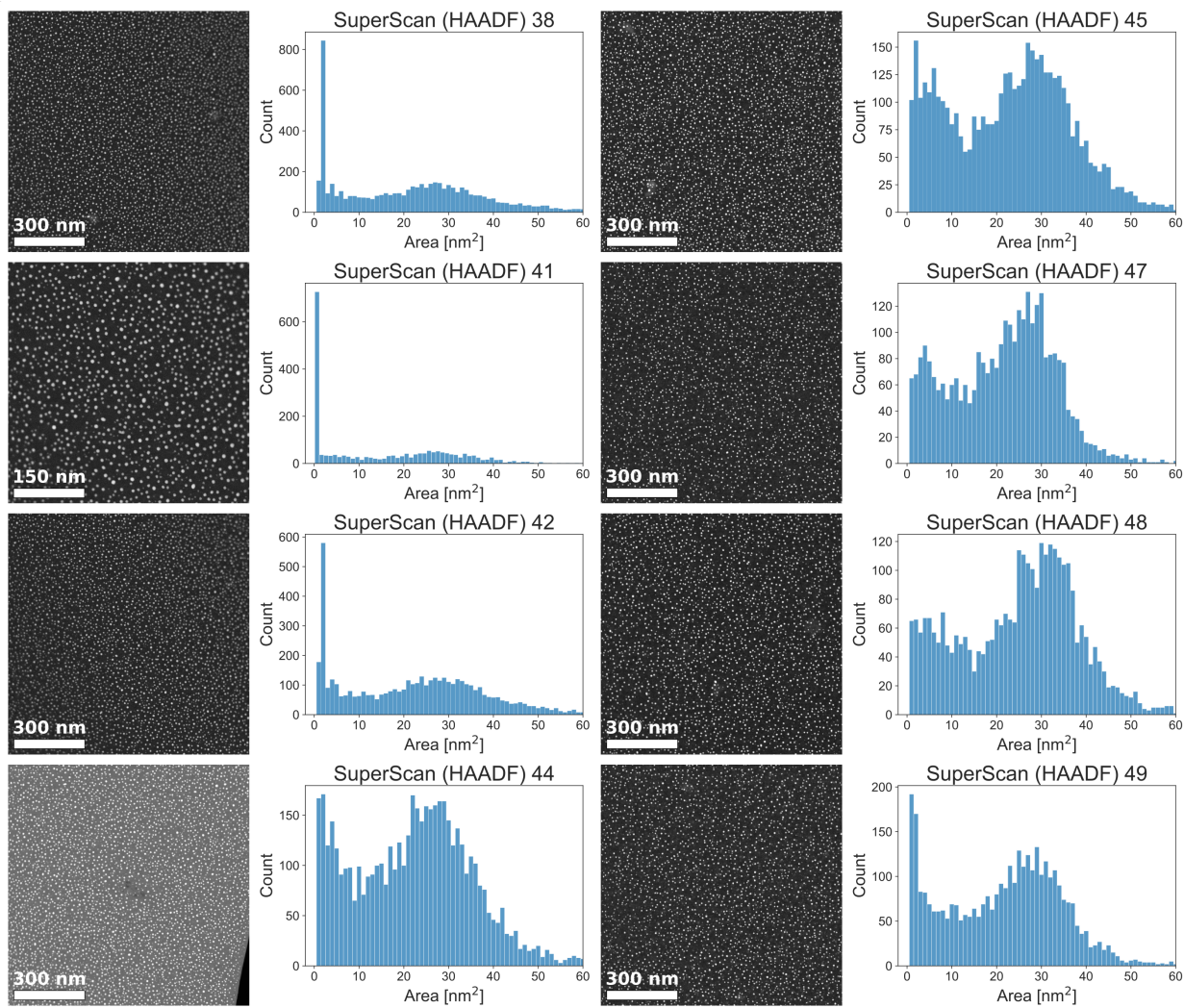

**Figure S1** Summary of nanoparticle size distribution for each image analyzed.

Electron energy loss spectroscopy (EELS) was recorded from the nanoplatelets. However, due to their sensitivity to e-beam irradiation recording the EELS signal from pristine nanoplatelets was not possible. The EELS signal shown in Figure S2 was recorded on the remnants of a nanoplatelet after substantial beam damage and crystallographic restructuring had occurred. We observe a clear C and Sn signal corresponding to the graphene and evaporated Sn atoms. We do not observe a clear O signal, however, it is possible that O has been selectively ejected from the nanoplatelet by the e-beam, driving the observed crystallographic transformation. Therefore, it remains unclear whether the nanoplatelets are 2D tin or a 2D tin oxide.

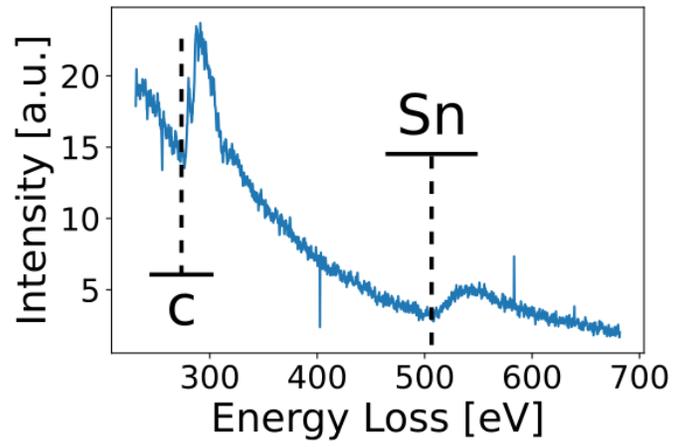

**Figure S2 EELS signal acquired on the remnant of a beam-damaged nanoplatelet.**